\documentclass[pre,aps,twocolumn, floatfix, superscriptaddress,showpacs]{revtex4}
\usepackage{amsmath,bm,graphicx}
\usepackage{amssymb}
\usepackage{amsfonts}
\renewcommand{\k}{{\mathbf k}}
\newcommand{\w}{\omega}

\newcommand{\N}[1]{N_{#1}}
\newcommand{\pd}[2]{\frac{\partial #1}{\partial #2}}
\newcommand{\dd}[2]{\frac{d #1}{d #2}}

\begin{document}

\title{Dynamical Scaling and the Finite Capacity Anomaly in 3-Wave Turbulence}

\author{Colm Connaughton}
\email{connaughtonc@gmail.com}
\affiliation {Centre for Complexity Science, University of Warwick, Gibbet Hill
Road, Coventry CV4 7AL, UK}
\affiliation{Mathematics Institute, University of Warwick, Gibbet Hill
Road, Coventry CV4 7AL, UK}
\author{Alan C. Newell}
\email{anewell@math.arizona.edu}
\affiliation {Dept. of Mathematics, University of Arizona, 617 N. Santa Rita Ave, Tucson, AZ 85721, U.S.A.}
\date{\today}

\begin{abstract}
We present a systematic study of the dynamical scaling process leading to the 
establishment of the Kolmogorov--Zakharov (KZ) spectrum in weak 3-wave turbulence. 
In the finite capacity case, in which the transient spectrum reaches infinite
frequency in finite time, the dynamical scaling exponent is anomalous
in the sense that it cannot be determined from dimensional considerations.
As a consequence, the transient spectrum preceding the establishment
of the steady state is steeper than the KZ spectrum. Constant energy flux
is actually established from right to left in frequency space after the 
singularity of the transient solution. 
From arguments based on entropy production, a steeper transient
spectrum is heuristically plausible.
\end{abstract}

\pacs{47.35.-i, 82.20.-w,  94.05.Lk}
\maketitle

Wave turbulence \cite{ZLF92} concerns the statistical mechanics of systems 
containing large numbers of dispersive waves which interact conservatively. 
Such wave systems occur in a variety of contexts in nature and engineering. 
Commonly cited examples include gravity waves on the ocean surface, waves in 
Bose-Einstein condensates and magnetohydrodynamic waves in strongly magnetized 
plasmas. For a short review of applications see \cite{CNN03}. Often the waves
are driven by external forcing which supplies energy (or possibly another
conserved quantity such as wave--action) at a scale which
is widely separated from the characteristic scale of dissipation. Since
interactions among waves are conservative, turbulence results. That is to say,
the physics is dominated by the flux of energy between the
forcing scale and dissipation scale. This flux is  mediated by the wave 
interactions. Theoretically, wave turbulence is more tractable than its
hydrodynamic cousin. In the limit of weak nonlinearity,
the theory is asymptotically closed under relatively mild assumptions on the
initial wave statistics (see \cite{NNB01} for a review). Asymptotic closure,
resulting from the interplay of weak nonlinearity and dispersion, allows 
a consistent derivation of a kinetic equation describing the long
time asymptotics of the frequency-space wave action density, $N_\w(t)$ 
\footnote{We shall limit ourselves to isotropic, scale-invariant systems so 
that we can exchange the $\k$-space kinetic equation for its simpler $\w$-space
analogue.}, which
in turn, determines the leading order behaviour of all the higher order
cumulants in both Fourier and physical spaces. In this article, we limit our 
discussion to the so-called 3-wave kinetic 
equation (3WKE) describing cases where the dominant nonlinearity is quadratic
and the dispersion relation admits three-wave resonances.

The 3WKE is a nonlinear integro-differential equation. It
generally involves 3 scaling exponents, traditionally written as $\alpha$, 
$\beta$ and $d$ representing the degree homogeneity of the wave
dispersion relation, the degree of homogeneity of the wave-wave interaction
coefficient and the spatial dimension respectively. For
isotropic systems, it can be written \cite{CON2009} in a form  involving
only a single exponent $\lambda=(2\beta-\alpha)/\alpha$:
\begin{equation}
\label{eq-3WKE}
\pd{\N{\w_1}}{t} = S_1[\N{\w}] + S_2[\N{\w}] + S_3[\N{\w}]
\end{equation}
where the collision integrals, $S_1[\N{\w}]$, $S_2[\N{\w}]$ and $S_1[\N{\w}]$, 
are written explicitly in the appendix. $S_1[\N{\w}]$ describes forward 
transfer of energy whereas $S_2[\N{\w}]$ and $S_1[\N{\w}]$ account for
backscatter. The details of the wave-wave interactions enter the collision
integrals through the wave interaction kernel, $K(\w_1,\w_2)$
\footnote{For the purposes of simplicity, we assume that $d=\alpha$ so that
there is no difference between the interaction kernels of $S_1[\N{\w}]$, $S_2[\N{\w}]$ and $S_1[\N{\w}]$. As a result the thermodynamic solution is $\w^{-1}$. 
See \cite{CON2009} for details.}. The exponent $\lambda$ is the degree of
homogeneity of this kernel. We shall focus particularly on
the product kernel:
\begin{equation}
\label{eq-productKernel}
K(\w_1,\w_2) = (\w_1\,\w_2)^\frac{\lambda}{2}.
\end{equation}
A key theoretical insight is the fact that Eq.~(\ref{eq-3WKE}) has a stationary
solution, the Kolmogorov--Zakharov (KZ) spectrum, 
$N_\w = c_{\rm KZ}\, \w^{-x_{\rm KZ}}$, which carries
a constant flux of energy through $\w$-space. The exponent, 
$x_{\rm KZ} = \frac{\lambda+3}{2}$, and the
constant, $c_{\rm KZ}$, can be found analytically using an elegant
technique known as the Zakharov transformation \cite{ZLF92}. The KZ spectrum
is the analogue of the Kolmogorov $5/3$ spectrum of hydrodynamic
turbulence. Considerable efforts have been made to realise
this spectrum experimentally. Extensive theoretical studies have 
completely characterised its locality and stability properties.

\begin{figure}[tbh]
\includegraphics[width=7cm]{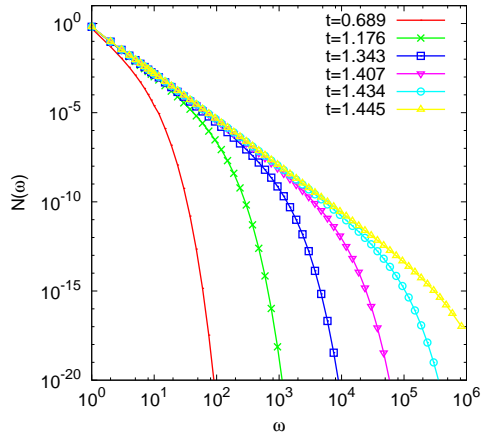}
\caption{Numerical solution of Eq.~(\ref{eq-3WKE}) with the $\lambda=2$ 
product kernel, Eq.~(\ref{eq-productKernel}), showing the singular transient 
dynamics preceding the establishment of the K-Z spectrum.  Numerics were 
performed using the algorithm described in \cite{CON2009}.}
\label{fig-timeEvolution}
\end{figure}

On the other hand, relatively little is known about the time-dependent
solutions of Eq.~(\ref{eq-3WKE}). This is the main topic of this article.  
Such solutions are important since they describe the process by which the KZ 
spectrum is established when an initially quiescent wave field is forced at 
large scales. Understanding the transient dynamics of kinetic equations like
Eq.~(\ref{eq-3WKE}) has become a problem of considerable importance in
understanding the non-equilibrium dynamics of Bose-Einstein condensation
\cite{LLPR2001,CP04} as well as in wave forecasting.  A basic scaling theory
of the transient dynamics was presented in \cite{FS1991}. Two important
observations were made in that paper. Firstly, the KZ spectrum is established
by the self-similar propagation of a front in $\w$-space which
moves towards large $\w$. Secondly, in finite capacity cases ($\lambda>1$) 
this front propagates to $\w=\infty$ in a finite 
time so that the KZ spectrum is set up by a singular solution of 
Eq.~(\ref{eq-3WKE}). A typical example is shown in Fig.~\ref{fig-timeEvolution}. When $\lambda \leq 1$ - the infinite capacity case -
the front necessarily leaves the KZ spectrum in its wake. For finite capacity
cases, the corresponding statement is more heuristic for reasons we will
examine below. The early numerical simulations in \cite{FS1991} indicated
that the transient spectrum had the KZ scaling. Subsequent numerical
investigations of finite capacity cascades, firstly in the context of plasma 
turbulence \cite{GNNP2000} and then in BEC \cite{LLPR2001} suggested that the 
transient scaling exponent, while very close to the KZ exponent, is not exactly
equal to it. This fact has become known as 
the ``finite capacity anomaly''. Investigations of finite capacity
cascades in differential equations \cite{CNP03,CN04} found that the
transient spectrum is always slightly steeper (for direct cascades) than the 
KZ spectrum, an observation which was heuristically justified on the basis of
entropy production considerations \cite{CNP03}.

In this article we perform a careful numerical study of the dynamical scaling
properties of the 3WKE and demonstrate that the finite capacity anomaly is
a generic feature of finite capacity cascades. The transient
spectrum is indeed steeper, although by a very small amount. Such a 
demonstration has been hitherto absent from the literature because of the
serious numerical difficulties encountered in solving Eq.~(\ref{eq-3WKE}) 
over a sufficient range of scales to measure exponents with sufficient
accuracy to demonstrate the anomaly. We procede as follows. We first 
introduce the idea of dynamical scaling, define the dynamical
scaling exponent, $a$,  and show how it relates to the transient spectrum. We 
then show that in the infinite capacity case, $a$ is given by the KZ value
and identify the failure in the corresponding argument in the finite capacity
case. We then turn to the delicate issue of numerical measurement of dynamical
scaling exponents which demonstrates that the transient spectrum is steeper
than the KZ spectrum in this case. We explore how the anomalous exponents
depend on the structure of the turbulence by varying the amount of backscatter
and the degree of scale--locality of the wave interactions. Finally we
show that a steeper transient spectrum is plausible on the basis of a heuristic
entropy production argument.

$N_\w(t)$ tends to a scaling (self-similar) form if there 
exists a monotonically increasing typical frequency, $s(t)$, a function, $F(x)$ 
and a dynamical scaling exponent, $a$, such that
\begin{equation}
\label{eq-scaling}
N_\w(t) \sim s(t)^a\,F\left(\frac{\w}{s(t)}\right).
\end{equation}
Here $\sim$ denotes the scaling limit: $s(t)\to\infty$ and $\w\to \infty$ with
$x = \w/s(t)$ fixed. The scaling function, $F(x)$, typically decays 
exponentially for large $x$ producing the front structure evident in 
Fig.~\ref{fig-timeEvolution}. The small $x$ behaviour of $F(x)$ determines
the transient spectrum. The general properties of the system are well 
characterised once the exponent $a$ and the small $x$ behaviour of $F(x)$
are known. We do not know, a--priori, that $F(x)$ is behaves as a power
near zero. Indeed we already know of one example of decaying wave turbulence 
in which $F(x)$ diverges as $x^{-1}\,\log(1/x)$ \cite{CK2009}. Nevertheless,
if we assume power law behaviour, $F(x) \sim x^{-y}$ as $x\to 0$, and further 
require that the spectrum should remain finite at the low frequency end as 
$s(t)\to \infty$, a reasonable assumption in the forced case, then we are
lead to conclude that $y=-a$ in order to cancel the time dependence. Therefore,
all we are required to determine is a single exponent, $a$.

Substitution of Eq.~(\ref{eq-scaling}) into Eq.~(\ref{eq-3WKE}) shows that
the typical frequency, $s(t)$, must evolve according to the equation:
\begin{equation}
\label{eq-sEvolution}
\dd{s}{t} = s^{\xi}\hspace{1.0cm}\mbox{with $\xi=\lambda + a +2$}
\end{equation}
while $F(x)$ must
satisfy the integro-differential equation:
\begin{equation}
\label{eq-nlinEvalProblem}
a F(x) + x\,\dd{F}{x} = S_1[F(x)] + S_2[F(x)] + S_3[F(x)].
\end{equation}
Three distinct behaviours for $s(t)$ arise depending on the value of
the exponent $\xi$: $s(t)$ grows algebraically in time (if $\xi<1$),
$s(t)$ grows exponentially in time (if $\xi=1$) or $s(t)$ exhibits a
finite time singularity of the form $(t^* - t)^{-\frac{1}{\xi-1}}$ (if
$\xi>1$). To determine what actually happens, we need to determine $a$.
For forced turbulence, the total energy grows linearly in time: 
$\int_0^\infty \w\,N_\w(t)\,d\w = J\,t$. Substituting the scaling form, 
Eq.~(\ref{eq-scaling}), into this equation and differentiating with respect to
time yields
\begin{equation}
\label{eq-energyGrowth}
\dd{s}{t} = \frac{J}{(a+2)\int_0^\infty x\,F(x)\,dx}\,s^{-1-a}.
\end{equation}
Comparing with Eq.~(\ref{eq-sEvolution}) fixes $a=-\frac{\lambda+3}{2}$, the 
KZ value. At first glance, we have shown that the dynamical scaling exponent
always takes its KZ value. Care is required however: 
Eq.~(\ref{eq-energyGrowth}) only holds provided the integral $\int x\,F(x)\,dx$
does not diverge at its lower limit on the predicted small $x$ behaviour, 
$F(x) \sim x^{-\frac{\lambda+3}{2}}$. This is only assured for $\lambda<1$ - 
the infinite capacity case. In the finite capacity case, conservation of
energy does not constrain the scaling function since the first moment of
$F(x)$ diverges. Physically, this is not mysterious: the scaling solution
Eq.~(\ref{eq-scaling}) does not probe the energy--containing scales in the
finite capacity case. As a result, the finite capacity case exhibits what 
Barenblatt refers to as self--similarity of the second kind \cite{BAR1996}: 
$a$ should be determined by solving Eq.~(\ref{eq-nlinEvalProblem}).

\begin{figure}[tbh]
\includegraphics[width=7cm]{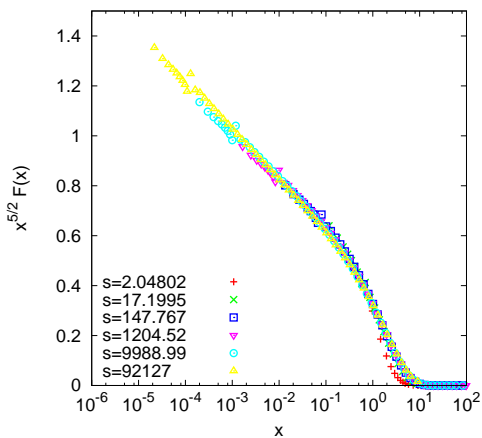}
\caption{Collapse of the $\frac{5}{2}$ order moment of the data in 
Fig.~\ref{fig-timeEvolution} according to the scaling hypothesis, 
Eq.~(\ref{eq-scaling}). The dynamical scaling exponent is $2.578 \pm 0.021$.}
\label{fig-dataCollapseKEd}
\end{figure}

There is no a--priori reason why $a$ should {\em not} be given by its KZ value
in the finite capacity case. Indeed one might naively hope it might
be based on dimensional analysis of Eq.~(\ref{eq-energyGrowth}). There is,
however an established precedent to indicate that it probably is not. The
kinetics of irreversible cluster-cluster aggregation (see \cite{LEY2003} for
a review), described by the Smoluchowski kinetic equation, has many structural 
similarities to 3-wave turbulence, including KZ spectra \cite{CRZ2004} although
it does not have an analogue of the thermodynamic spectrum. In fact,
the Smoluchowski equation simply corresponds to Eq.~(\ref{eq-3WKE}) with the
backscatter terms, $S_2(N_\w)$ and $S_3(N_\w)$ removed. The non-triviality of
dynamical scaling exponents is well known \cite{CS1997} in that field. An
extensive numerical study performed by Lee \cite{LEE2001} convincingly
demonstrated the existence of the finite capacity anomaly for the Smoluchowski
equation and showed, intriguingly, that it has the opposite sign to the
putative anomaly in the 3WKE: the transient spectrum in the Smoluchowski
equation is shallower than the KZ spectrum. Our method extends Lee's approach
to include the back--scatter terms so we obviously reproduce Lee's results and
various known exact solutions of the Smoluchowski equation when we turn
the back--scatter terms off.

We now turn to the direct measurement of the dynamical exponent from numerical
data in order to demonstrate the anomaly in the 3WKE. The details of the
numerical algorithm and its validation are described in \cite{CON2009}. There are two principal
challenges. The first is the determination of the typical scale, $s(t)$.
The second is the determination of the value of $a$ which provides the
best data collapse according to Eq.~(\ref{eq-scaling}). Let us now address
these challenges in turn. The typical scale
can be determined intrinsically by measuring moments of the wave spectrum,
$M_\sigma (t) = \int_0^\infty \w^\sigma\,N_\w(t)\,d\w$. Eq.~(\ref{eq-scaling})
implies that $s(t)$ is given by the ratio of successive moments: 
$s(t) = M_{\sigma+1}(t)/M_\sigma(t)$. We have already learned to
be wary of divergences of moments in the scaling limit coming from the 
behaviour of $F(x)$ near 0. This issue arises again in the definition of
$s(t)$. We must take a ratio of successive sufficiently high order moments in order
to ensure the convergence of the necessary integrals. In this paper, we
take $s(t) = M_3(t)/M_2(t)$ and restrict ourselves to 
$0\leq \lambda \leq 2$ which avoids this pitfall. The challenge in determining
$a$ from the numerical data lies in the fact that it is very close to the
KZ value so a very robust and sensitive data analysis is required to 
measure it with sufficient accuracy. For example, the simulation shown
in Fig.~\ref{fig-timeEvolution} has $a=2.578 \pm 0.021$ compared to
a KZ value of $2.5$. It is our experience that a "best-by-eye" measurement
of the slope of the wake in a log-log plot such as Fig.~\ref{fig-timeEvolution}
is completely inadequate. We have instead opted to exploit 
Eq.~(\ref{eq-scaling}) in its entirety and try to find the value of $a$
which best collapses all the curves in Fig.~\ref{fig-timeEvolution} onto
a single curve. This was done, as suggested in \cite{BS2001} by defining
a function which measures the total average spread of the collapsed data
for a given value of $a$ and then performing a one--dimensional minimization
of this function over $a$. To ensure maximum sensitivity of the analysis,
the data collapse was actually performed on the logartithm of 
$(\lambda+3)/2$-moment of $N_\w(t)$. In addition from removing the subjectivity
from the data collapse process, the accuracy of the resulting exponent can
be estimated from the width of the minimum of the data spread function. We 
chose to measure the width of the minimum at the 1\% level set of the data 
spread function.  A representative sample of the data collapse obtained by this
method is shown in  Fig.~\ref{fig-dataCollapseKEd}. It is clearly evident
that the data collapse is of a very high quality and that $F(x)$ 
is steeper than $x^{-5/2}$, the KZ value for this particular choice of kernel.

\begin{figure}[tbh]
\includegraphics[width=7cm]{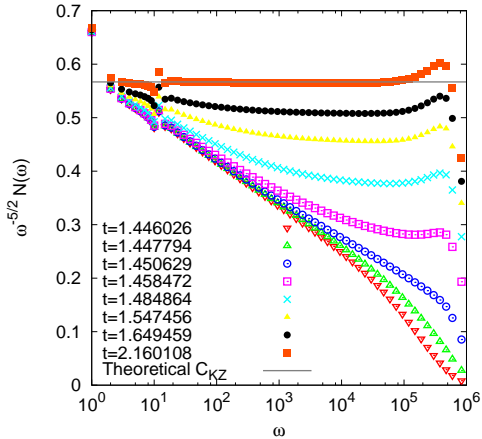}
\caption{Relaxation of the transient spectrum to the KZ spectrum 
post-singularity.}
\label{fig-postSingularity}
\end{figure}

\begin{figure}[tbh]
\includegraphics[width=7cm]{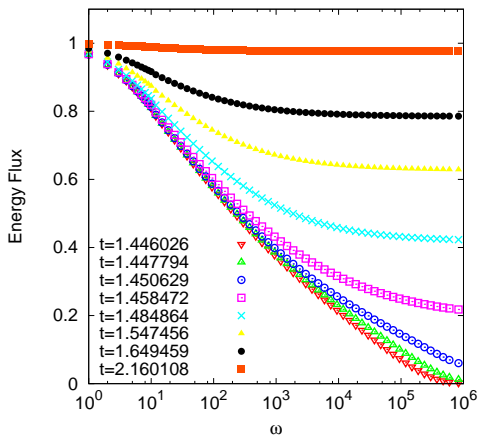}
\caption{Development of a constant energy flux post-singularity.}
\label{fig-energyFlux}
\end{figure}

In the numerical simulation, the singularity is regularised by the onset of
dissipation \cite{CON2009} which allows us to study the subsequent 
establishment of the KZ spectrum after the transient spectrum has reached
the dissipation scale. This is illustrated in Fig.~\ref{fig-postSingularity} 
which shows the relaxation of the compensated spectrum to the KZ-scaling. 
Note that the KZ spectrum is actually set up from right to left. That is to
say, the KZ spectrum first emerges at high $\w$ and then propagates to low $\w$ 
as a ``backwards'' front. This interpretation of the post-singularity 
dynamics is supported by measurements of the energy flux such as those
presented in Fig.~\ref{fig-energyFlux}. We remark that the energy flux
becomes flat as a function of $\w$ first at large $\w$ and this region
then spreads backwards. The formation of the KZ spectrum from the right
is consistent with previous simulations of the MHD \cite{GNNP2000} and
differential \cite{CNP03} kinetic equations although the present numerical
scheme is far more reliable. This process happens very quickly - notice the
timings of the successive snapshots in Figs.~\ref{fig-postSingularity}
and \ref{fig-energyFlux}. 

\begin{figure}[tbh]
\includegraphics[width=7cm]{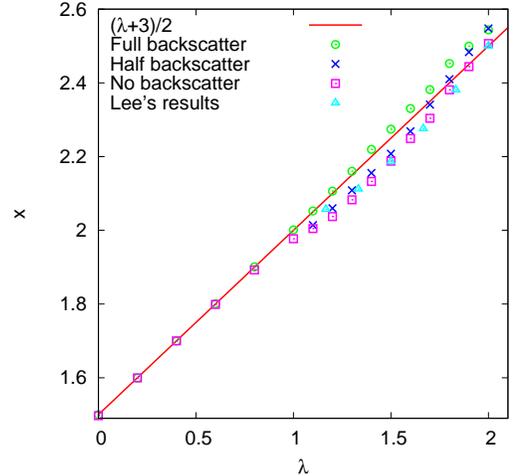}
\caption{Dynamical scaling exponent as a function of $\lambda$ for different amounts of backscatter. Lee's results for the Smoluchowski equation \cite{LEE2001} are shown for comparison.}
\label{fig-exponentsVaryBS}
\end{figure}

\begin{figure}[tbh]
\includegraphics[width=7cm]{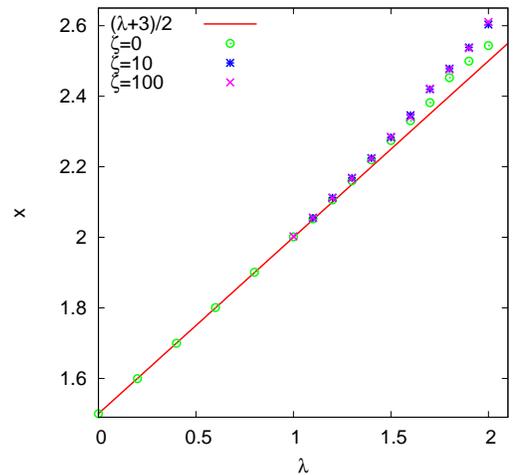}
\caption{Dynamical scaling exponent as a function of $\lambda$ for different degrees of locality.}
\label{fig-exponentsVaryLocality}
\end{figure}

When it comes to understanding the meaning of the anomalous scaling exponents,
one may take a mathematical point of view and simply view them as being 
nothing more than the exponents defined by the solution of the nonlinear 
eigenvalue problem Eq.~(\ref{eq-nlinEvalProblem}). From a physical perspective
this, while clearly a correct explanation, is a rather unsatisfactory one and
one cannot help exploring how the anomaly depends on the physical parameters
of the problem in the hope of obtaining a deeper insight. We performed
several systematic studies. Firstly, and probably most importantly, we measured
the dynamical exponent for a range values of $\lambda$ between 0 and 2 taking
the product kernel, Eq.~(\ref{eq-productKernel}) as a benchmark case. The
results the circular data points plotted in 
Fig.~\ref{fig-exponentsVaryBS} along with the KZ exponents. They demonstrate 
that the anomaly is generic for the finite capacity regime ($\lambda>1$) and 
increases with $\lambda$ although it is always a small correction to the KZ
value. From a certain point of view, the smallness of the anomalous correction
to the dynamical exponent is one of its most mysterious features. The 
other sets of data plotted on Fig.~\ref{fig-exponentsVaryBS} show how the
anomalous exponents vary with the amount of backscatter. This was done
by repeating the simulations, first  with the $S_1$ and $S_2$ terms in 
Eq.~(\ref{eq-3WKE}) reduced by a factor of 2 and then with them removed 
entirely (Smoluchowski equation). While it is clear that the sign of the
correction can be changed in this way, we stop short of offering a suggestion
of what this actually means. In any case, we are not at liberty to tune
the structure of the kinetic equation in this way in any conceivable
experiment. A second exploration, presented in 
Fig.~\ref{fig-exponentsVaryLocality}, probed how the anomalous exponent
varies with the degree of scale-locality of the wave interactions. This
was done by introducing a deformation of the interaction kernel, parameterised
by $\zeta$, which suppresses interactions between waves of widely different
frequencies while leaving the overall degree of homogeneity, $\lambda$,
 unchanged:
\begin{equation}
\label{eq-localisedProductKernel}
K_\zeta(\w_1,\w_2) = K(\w_1,\w_2)\,\exp\left[-\zeta\left(\frac{\w_1\w_2}{(\w_1+\w_2)^2}-\frac{1}{4}\right)\right].
\end{equation}
The results show that the anomaly gets bigger the more the non-local 
interactions are suppressed. This was a surprise for us but is
consistent with the fact that the anomalies found  in differential
approximations \cite{CNP03,CN04} were somewhat larger.

\begin{figure}[tbh]
\includegraphics[width=7cm]{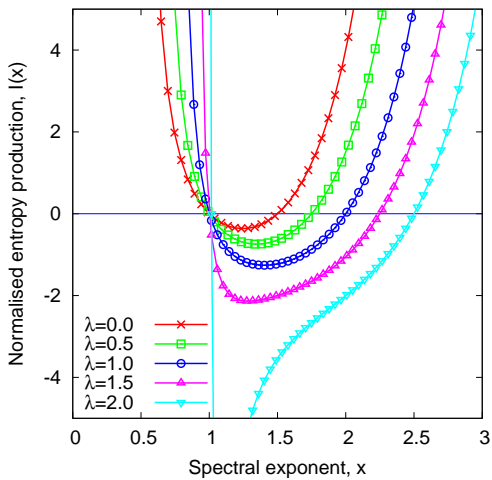}
\caption{Sign of the total entropy production on power law spectra, $N_\w=c\,\w^{-x}$ for the product kernel, Eq.~(\ref{eq-productKernel}), for a range of values of $\lambda$. 
For this kernel, the integral $I(x)$ given by Eq.~(\ref{eq-I}) is 
convergent for $x \in \left(\frac{\lambda}{2},3+\frac{\lambda}{2}\right)$ \cite{CON2009}.}
\label{fig-entropyProduction}
\end{figure}

Finally we offer a heuristic physical argument why a steeper spectrum
might be natural for the 3WKE. Let us formally define an entropy-like
quantity, $S_\w(t) = \log(N_\w(t))$. In the absence of small scale
regularization, $S_\w(t)$ diverges on typical realisations of $N_\w(t)$.
Nevertheless, its rate of production formally makes sense if it is
defined via the righthand side of Eq.~(\ref{eq-3WKE}): 
$\pd{S_\w(t)}{t} = \frac{1}{N_\w(t)}\,\pd{N_\w(t)}{t}$. One may then ask
the question of whether the entropy production is positive or
negative on a (not-necessarily stationary) power-law spectrum, 
$N_\w = c\,\w^{-x}$. Through application of the Zakharov transformation
one finds:
\begin{equation}
\pd{S_\w(t)}{t} = c\, \w^{2\,x_{\rm KZ} - x -2} I(x) 
\end{equation}
where
\begin{eqnarray}
\nonumber I(x) &=& \int_0^1 K(y,1-y) (y (1 - y))^{-x} ( 1 - y^x - (1-y)^x)\\
\label{eq-I}& & (1 - y^{2 x -\lambda - 2} - (1 - y)^{2 x -\lambda - 2}).
\end{eqnarray}
Whether the entropy production is positive or negative for a given 
spectral exponent, $x$, depends on the sign of $I(x)$. This integral 
is plotted as a function of $x$ for the product kernel with several values of 
$\lambda$ in Fig.~\ref{fig-entropyProduction}. Each curve has two zeros,
signifying vanishing entropy production (although vanishing for different 
reasons). They correspond to the thermodynamic
($x=1$) and KZ exponents. The entropy production is always negative in
between the two. Given that the transient spectrum is still exploring its
phase space, one would not expect the entropy production to be negative. If
we therefore limit transient spectra to those with positive entropy
production, Fig.~\ref{fig-entropyProduction} clearly requires any
candidate exponents to be greater than the KZ exponent. From this point of
view, the finite and infinite capacity cases are quite different in the sense
that for infinite capacity systems no entropy production occurs in the wake
and the phase space exploration occurs entirely at the front whereas in 
the finite capacity case it takes place throughout the full range of
scales.

To summarise, we have used what is probably the most accurate numerical
scheme currently available for solving the 3WKE to demonstrate the presence
of a small but definite finite capacity dynamical scaling anomaly for a 
set of kinetic equations with model wave interaction kernels given by
Eq.~(\ref{eq-productKernel}) and Eq.~(\ref{eq-localisedProductKernel}).
The results are completely consistent with previous observations from the 
original work on the Alfven wave turbulence to the differential equation models
suitable for ultra local transfer. The natural questions arise: is the 
anomalous realization of the KZ spectrum seen here the general property of all 
finite capacity situations such as, for example, 3D high Reynolds number 
hydrodynamic turbulence? If so, is there a general entropy production principle 
which is responsible for guiding the system towards the statistically steady 
state in this anomalous manner?

\appendix
\section{Collision integrals and the wave interaction kernel}
The explicit forms of the collision integrals which appear in 
Eq.~(\ref{eq-3WKE}) are given below. A full derivation appears in 
\cite{CON2009}.
\begin{eqnarray}
\nonumber S_1[\N{\w}]\!\!\! &=& \!\!\!\!\!\int\!\! K_1(\w_3,\w_2)\, \N{\w_2} \N{\w_3} \delta(\w_1\!-\!\w_2\!-\!\w_3)\, d\w_{23}\\
\label{eq-S1} &-& \!\!\!\!\!\int\!\! K_1(\w_3,\w_1)\, \N{\w_1} \N{\w_3} \delta(\w_2\!-\!\w_3\!-\!\w_1)\, d\w_{23}\\
\nonumber  &-& \!\!\!\!\!\int\!\! K_1(\w_1,\w_2)\,\N{\w_1} \N{\w_2} \delta(\w_3\!-\!\w_1\!-\!\w_2)\, d\w_{23},
\end{eqnarray}
\begin{eqnarray}
\nonumber S_2[\N{\w}]=\!\!\! &-& \!\!\!\!\!\int\!\! K_2(\w_3,\w_2)\, \N{\w_1} \N{\w_2} \delta(\w_1\!-\!\w_2\!-\!\w_3)\, d\w_{23}\\
\label{eq-S2} &+& \!\!\!\!\!\int\!\! K_2(\w_3,\w_1)\, \N{\w_2} \N{\w_3} \delta(\w_2\!-\!\w_3\!-\!\w_1)\, d\w_{23}\\
\nonumber  &+& \!\!\!\!\!\int\!\! K_2(\w_1,\w_2)\,\N{\w_1} \N{\w_3} \delta(\w_3\!-\!\w_1\!-\!\w_2)\, d\w_{23}
\end{eqnarray}
and
\begin{eqnarray}
\nonumber S_3[\N{\w}]=\!\!\! &-& \!\!\!\!\!\int\!\! K_3(\w_3,\w_2)\, \N{\w_1} \N{\w_3} \delta(\w_1\!-\!\w_2\!-\!\w_3)\, d\w_{23}\\
\label{eq-S3} &+& \!\!\!\!\!\int\!\! K_3(\w_3,\w_1)\, \N{\w_1} \N{\w_2} \delta(\w_2\!-\!\w_3\!-\!\w_1)\, d\w_{23}\\
\nonumber  &+& \!\!\!\!\!\int\!\! K_3(\w_1,\w_2)\,\N{\w_2} \N{\w_3} \delta(\w_3\!-\!\w_1\!-\!\w_2)\, d\w_{23}.
\end{eqnarray}
The wave interaction kernels, $K_i(\w_1,\w_2)$ ($i=1,2,3$), all have degree of
homogeneity $\lambda=(2\beta-\alpha)/\alpha$.
There is a price to be paid for removing all explicit dependence on the spatial 
dimension from the kinetic equation. The kernels
appearing in the forward transfer integral and the backwards transfer 
integrals are not, in general, the same.
Furthermore, they are not symmetric in their arguments as is the case when
the kinetic equation is written in its  usual form (as in \cite{ZLF92} or 
\cite{NNB01} for example).
The relationship between them is, however, straightforward:
\begin{eqnarray}
\nonumber K_2(\w_i,\w_j) &=& K_1(\w_i,\w_j) \left(\frac{\w_i+\w_j}{\w_j}\right)^{\frac{\alpha-d}{\alpha}}\\
\label{eq-defnK2K3}K_3(\w_i,\w_j) &=& K_1(\w_i,\w_j) \left(\frac{\w_i+\w_j}{\w_i}\right)^{\frac{\alpha-d}{\alpha}}.
\end{eqnarray}
Each of the collision integrals, taken individually has a finite flux stationary
solution, $N_\w = c_{KZ}\,\sqrt{J}\,\w^{-\frac{\lambda+3}{x}}$, where $J$
is the energy flux and $c_{KZ}$ is a constant. This may be demonstrated by
applying the appropriate Zakharov transformations \cite{ZLF92} to each 
integral in turn. The stationary thermodynamic solution is hidden in this
representation. It is recovered by recombining all three collision integrals 
and using the relationships between the collision kernels. One obtains the
thermodynamic solution $N_\w = c_T\,w^{-1-\frac{\alpha-d}{\alpha}}$ where 
$c_T$ is constant. In the present work, for simplicity, we take $\alpha=d$ so 
that the distinctions between the forward and backward interactions disappear 
and the thermodynamic solution is simply $\w^{-1}$.
\end{document}